# *Terms-we-Serve-with*: a feminist-inspired social imaginary for improved transparency in algorithmic decision-making

Bogdana Rakova*, Mozilla Foundation, Senior Trustworthy AI Fellow
Megan Ma*, Stanford Law School, CodeX Fellow
Renee Shelby*, Google, Senior Responsible Innovation Researcher
* All authors are equal contributors

> *"Power flows through governing bodies, social institutions, and micro-interactions, all of which engage with technologies of the time."*
> *- Jenny Davis, Apryl Williams, and Michael W Yang* [1]

## Abstract

Power and information asymmetries between people and digital technology companies have predominantly been legitimized through contractual agreements that have failed to provide diverse people with meaningful consent and contestability.[2] We offer an interdisciplinary multidimensional perspective on the future of regulatory frameworks - the Terms-we-Serve-with (TwSw) social, computational, and legal contract for restructuring power asymmetries and center-periphery dynamics to enable improved human agency in individual and collective experiences of algorithmic harms.[3]

**Co-constitution**
Experiences of algorithmic harms are shaped by contractual agreements. We challenge coercive terms-of-service through multi-stakeholder engagement. Marginalized communities and various stakeholders take part in drafting TwSw agreements and are rewarded for their participation. We believe they can help technology companies anticipate algorithmic harm and design adequate response mechanisms that empower solidarity.[4]

**Productive Friction**
TwSw agreements acknowledge and reflect marginalized knowledge systems. We challenge one-sided terms-of-service by enabling meaningful dialogue through the production and resolution of conflict. We believe critical discussion allows individuals to self-organize and discuss algorithmic harms and accountability mechanisms in a way that is safe, respects their privacy, and human dignity.

---

[1] Jenny Davis, Apryl Williams, and Michael W Yang. 2021. Algorithmic reparation. Big Data & Society, 8(2), 20539517211044808.
[2] Ewa Luger, Stuart Moran, and Tom Rodden. 2013. Consent for all: revealing the hidden complexity of terms and conditions. In Proceed- 1194 ings of the SIGCHI conference on Human factors in computing systems. 1195 2687–2696. Jonathan A Obar and Anne Oeldorf-Hirsch. 2020. The biggest lie on the internet: Ignoring the privacy policies and terms of service policies of social networking services. Information, Communication & Society 23, 1 (2020), 128–147. Jacob Metcalf, Emanuel Moss, Elizabeth Anne Watkins, Ranjit Singh, and Madeleine Clare Elish. 2021. "Algorithmic Impact Assessments and Accountability: The Co-Construction of Impacts," 12.
[3] We define *algorithmic harm* as incidents experienced by individuals and communities that lead to social, material, or ecological harms, resulting from algorithmic systems and interactions between human and algorithmic actors.
[4] Bogdana Rakova, Jingying Yang, Henriette Cramer, and Rumman Chowdhury. "Where responsible AI meets reality: Practitioner perspectives on enablers for shifting organizational practices." Proceedings of the ACM on Human-Computer Interaction 5, no. CSCW1 (2021): 1-23.

**Veto Power**
We recognize that people hold a multiplicity of dynamic human identities. Inspired by the temporal dynamics in law - writing in the present, reflecting on the past, and encoding the future, we challenge the dominant temporal design of algorithmic systems which have been critically framed as self-fulfilling prophecies[5] and stochastic parrots.[6] In contrast, TwSw reflect the temporal dynamics of how individual and collective experiences of algorithmic harm unfold and include socio-technical mechanisms that facilitate veto power.

**Verification**
We believe individuals and communities could build and leverage open-source tools in verifying that TwSw are being met. We point to the use of computable contracts as a mechanism to verify properties of algorithmic outcomes and enable reporting of algorithmic harm. We see this as a new feedback mechanism between technology companies and civil society stakeholders who have the expertise to take action in helping individuals and communities.

**Accountability**
We acknowledge that reporting and the perceptions it carries are complex. With TwSw, we imagine that users can hold technology companies accountable by leveraging dispute resolution mechanisms with the notion of apology at heart. Seizing traditional courts is rather arduous with the complexity and opacity of formal legal processes frequently leading to injury. By contrast, mediation is an alternative form of dispute resolution that circumvents adversarial contexts, and instead, centers on reconciliation through apology.

Apology has played a significant role in existing fields of law, particularly in circumstances of medical error.[7] Expressions of regret recognize imperfection and the space for change. Similarly, we see TwSw as embodying an analogous enforcement mechanism; one that can enable an alternative apology-centered mediation. We imagine a Forgiveness Forum inspired by reparative algorithms[8] that unmask and undo algorithmic harm, but extend further to unpack harm through apology.

In conclusion, we hope that our design perspective for the Terms-we-Serve-with agreement could inspire new forms of regulatory frameworks that take into account how considering a multiplicity of possible futures could lead to transformative change in the present.

# Appendix

**Terms-we-Serve-with, Algorithmic Harms, and Reparation**
"I agree to the terms of service" is perhaps the most falsely given form of consent,[9] often leaving individuals powerless in cases of algorithmic harm - incidents experienced by individuals and communities that lead to social, material, or ecological harms, resulting from algorithmic systems and interactions between human and algorithmic actors. Our TwSw agreement manifesto is meant to enable various stakeholders to better anticipate and foster accountability when algorithmic harms arise.

---

[5] Pamela Ugwudike. 2021. Data-Driven Algorithms in Criminal Justice: Predictions as Self-fulfilling Prophecies. In Kohl, U. and Eisler, J. (eds.), Data-Driven Personalisation in Markets, Politics and Law, chapter, Cambridge, Cambridge University Press, pp. 190–204.
[6] Emily M. Bender, Timnit Gebru, Angelina McMillan-Major, and Shmargaret Shmitchell. On the Dangers of Stochastic Parrots: Can Language Models Be Too Big? . In Proceedings of the 2021 ACM Conference on Fairness, Accountability, and Transparency, pp. 610-623. 2021.
[7] Apologies and Medical Error, https://www.ncbi.nlm.nih.gov/pmc/articles/PMC2628492/
[8] *Id at 1.*
[9] Ewa Luger, Stuart Moran, and Tom Rodden. 2013. Consent for all: revealing the hidden complexity of terms and conditions. In Proceed- 1194 ings of the SIGCHI conference on Human factors in computing systems. 1195 2687–2696. Jonathan A Obar and Anne Oeldorf-Hirsch. 2020. The biggest lie on the internet: Ignoring the privacy policies and terms of service policies of social networking services. Information, Communication & Society 23, 1 (2020), 128–147.

The TwSw dimensions we have laid out are in conversation with other feminist efforts to transform the power relations in algorithmic systems. The [Feminist Principles of the Internet](#) use the lens of gender and sexuality rights to charter seventeen 'critical internet-related rights' in terms of access, movements, economy, expression, and embodiment. [Traditional Knowledge Labels](#) enable local, community control over the access and use of indigenous knowledge. The [Feminist Data Manifest-No](#) offers refusals and commitments to create 'new data futures' where data-driven harms are minimized through community control over data knowledges. The [Design Justice Network](#) collectively developed principles to rethink design processes to center those traditionally marginalized by design practices to 'sustain, heal, and empower communities.' The [carceral tech resistance network](#) supports community-led research, archiving and database building, and training to educate about carceral technologies that put communities at risk. [Our Data Bodies](#) developed the [Digital Defense Playbook](#) which offers activities to co-create knowledge and tools for communities working in the 'intersectional fight for racial justice, LGBQT liberation, feminism, immigrant rights, economic justice and other freedom struggles' to co-create and share knowledge to understand and address the impact of data-centric systems. Building on their work, the TwSw are in solidarity with these transformational efforts to build futures where the power dynamics that foster algorithmic harms are dismantled.

**The Terms-we-Serve-with Dimensions**
In computing, the so-called *principle component analysis,*[10] commonly used for dimensionality reduction, is a method for increasing interpretability through identifying dimensions (principal components) of complex data in a way that preserves the most information. The TwSw dimensions — co-constitution, productive friction, veto power, verification, and accountability — are methods for cultivating and preserving critical knowledge and relations. We hope to leverage these dimensions in service of algorithmic harms-reduction and to co-create algorithmic systems that empower racialized women, non-binary people, marginalized groups in the Global South, and others who have historically been misrepresented in the development of algorithmic systems.[11] We also recognize there can be infinite dimensions, and hold space for new TwSw dimensions to emerge.

**Speculative Case Study**
Recently, a range of critiques, audits, and empirical analysis of algorithmic systems has drawn attention to the environmental harms arising from their design, development, and deployment. Interdisciplinary scholars have investigated the carbon footprint of training large scale machine learning models,[12] activists and grassroots communities have called for the need to bring awareness about the use of algorithmic systems in industries such as fossil fuels and mining operations,[13] while there has also been a small number of resistance efforts lead by employees of technology companies including Amazon[14] and Microsoft.[15]

---

[10] Xiangyu Kong, Changhua Hu, and Zhansheng Duan. Principal component analysis networks and algorithms. Springer Singapore, 2017.
[11] Renee Shelby. 2021. Technology, sexual violence, and power-evasive politics: Mapping the anti-violence sociotechnical imaginary. Science, Technology, & Human Values, 01622439211046047; Ricaurte, Paola. 2019. "Data epistemologies, the coloniality of power, and resistance." Television & New Media 20, no. 4: 350-365.
[12] *Id at 5.;* Schwartz, R., J. Dodge, N. A. Smith, and O. Etzioni, 2020: Green AI. Communications of the ACM, 63 (12), 54–63. Xu, J., W. Zhou, Z. Fu, H. Zhou, and L. Li, 2021: A survey on green deep learning. arXiv preprint arXiv:2111.05193.
[13] Zero Cool. 2019. Oil is the New Data. Logic. Issue 9, "Nature". Available at: [https://logicmag.io/nature/oil-is-the-new-data/](https://logicmag.io/nature/oil-is-the-new-data/); Shannon Dosemagen, Evelin Heidel, Luis Felipe R. Murillo, Emilio Velis, Alex Stinson and Michelle Thorne. Open Climate Now!. Branch Magazine. Available at https://branch.climateaction.tech/issues/issue-2/open-climate-now/
[14] Amazon Employees for Climate Justice. 2019. "Amazon employees are joining the Global Climate Walkout, 9/20". Medium. https://amazonemployees4climatejustice.medium.com/amazon-employees-are-joining-the-global-climate-walkout-9-20-9bfa4cbb1ce3
[15] Microsoft Employees. 2019. Microsoft Workers for Climate Justice. GitHub. https://github.com/MSworkers/for.ClimateAction

A digital technology company **MetaForest** is developing advanced data science, machine learning, and drone automation to restore biodiverse ecosystems at scale. They've been concerned about the broader impact of their algorithmic systems and have decided to adopt the Terms-we-Serve-with Manifesto at the core of their business model.

- **Co-constitution** - MetaForest's interdisciplinary engineering team has built a deep understanding that data and knowledge are interconnected. Through participatory action research methods, they have engaged with historically misrepresented communities in the drafting of their Terms-we-Serve-with agreement. This was done through interactive workshops where community members were rewarded for the value and knowledge systems they hold. Complex issues were discussed, for example, with regards to Traditional Ecological Knowledge[16] in the situated context where MetaForest's technology is being deployed. Any future changes made to the Terms-we-Serve-with agreement need to be approved during participatory workshops with local communities, policymakers, and other stakeholders involved in the local context where the technology is deployed.
- **Productive Friction** - MetaForest has created meaningful mechanisms that enable historically misrepresented communities to be equal stakeholders in the development and impact assessment of their algorithmic systems.[17] For example, they've built open-source tools and invited external communities to engage in environmental data monitoring of the impact of the algorithmic system in the places where it's deployed. Working with the "productive frictions between emerging data justice concerns and long standing principles of environmental justice,"[18] MetaForest is partnering with the Environmental Data and Governance Initiative[19] and has utilized their syllabus[20] in internal employee training programs.
- **Veto Power** - As part of the lifecycle of an algorithmic system, MetaForest has considered its design, development, deployment, ongoing monitoring and maintenance, as well as its decommission. MetaForest has adopted a place-based[21] and relational approach[22] to the impact assessment of their algorithmic systems. Critically, their Terms-we-Serve-with agreement also governs in what cases will a machine learning model be decommissioned, for example, due to biases arising from the dynamic changes in the natural ecosystems where the technology is deployed, considering the ecosystem services and services to ecosystem within the specific local context.[23]
- **Verification** - A recommendation system (machine learning model) built by MetaForest utilizes satellite and drone image data to help local landowners sell carbon credits derived from their land. The credits correspond to the estimated amount of emissions which are saved through their efforts in sustaining a thriving ecosystem. The credits need to be validated by a regulatory mechanism before they can be sold by brokers in a carbon market. There is high risk that an algorithmic system used in this process might reproduce historical inequities and unfairly impact farmers in the Global South. The TwSw agreement could bring improved transparency to the credits verification algorithmic system by enabling a computable verification contract which executes automatically on individual algorithmic recommendations. Smart contracts are already used as a transaction mechanism within carbon markets. In contrast to a transaction mechanism, the goal of a computable contract is to provide compliance checking. In addition to computable contracts, the TwSw will enable a reporting mechanism which allows farmers to question equity and justice concerns they might have with unfair estimations performed by the algorithmic system.
- **Accountability** - as a radical approach to considering their impact on climate justice, MetaForest is building tools that enable local communities where their technology is deployed to uphold the emerging notion of environmental personhood for natural entities like rivers and mountains. Environmental personhood grants legal personhood rights to nature, providing the ability for nature to interface with the other stakeholders within legal systems,[24] through appointed guardians who are local indigenous peoples.[25] Furthermore, MetaTree's TwSw agreement pays special attention to the governance of data used by their algorithmic system. They've adopted the Kaitiakitanga affirmative action data license developed by a Maori community in New Zealand, stating that any benefits derived from local communities' data will flow back to the source of the data.[26]

---

[16] Fikret Berkes. 2017. Sacred Ecology: Traditional Ecological Knowledge and Resource Management. Routledge
[17] Hommels, A., Mesman, J. and Bijker, W.E. eds., 2014. Vulnerability in technological cultures: New directions in research and governance. MIT Press.
[18] Vera, Lourdes A., Dawn Walker, Michelle Murphy, Becky Mansfield, Ladan Mohamed Siad, Jessica Ogden, and EDGI. "When data justice and environmental justice meet: formulating a response to extractive logic through environmental data justice." Information, Communication & Society 22, no. 7 (2019): 1012-1028.
[19] *Id.*
[20] Environmental Data Justice Working Group. 2017. An Environmental Data Justice Syllabus 1.0. Available at: https://docs.google.com/document/d/1O7ytnzXWFkluiYE4Pulo_mCHs9jdNpPm8hw83aLU2pg/edit#heading=h.p7r8t8xelqcr
[21] de Vos, A., Biggs, R. and Preiser, R., 2019. Methods for understanding social-ecological systems: a review of place-based studies. *Ecology and Society*, 24(4).
[22] Birhane, A. and Cummins, F., 2019. Algorithmic injustices: Towards a relational ethics. *arXiv preprint arXiv:1912.07376*.
[23] Comberti, C., Thornton, T.F., de Echeverria, V.W. and Patterson, T., 2015. Ecosystem services or services to ecosystems? Valuing cultivation and reciprocal relationships between humans and ecosystems. Global Environmental Change, 34, pp.247-262.

We hope that the Terms-we-Serve-with proposal serves as an invitation for a pluriversal dialogue where we collectively engage in co-creating a new trustworthy social imaginary for improved transparency and human agency in the contractual agreements between people and algorithmic decision-making systems.

---